\documentclass[onecolumn,nobibnotes,nofootinbib]{revtex4}
\usepackage{epsfig}
\usepackage{amsmath,amssymb,bm}
\newcommand{\avg}[1]{\left\langle #1 \right\rangle}
\newcommand{\calN}{{\cal N}}
\newcommand{\rr}{\bm{r}}
\newcommand{\bb}{\bm{b}}
\newcommand{\xx}{\bm{x}}
\newcommand{\yy}{\bm{y}}
\newcommand{\zz}{\bm{z}}

\begin{document}
\title{Probing gluon number fluctuation effects in future electron-hadron colliders}
\author{J. T. Amaral ${}^1$, V.P. Gon\c{c}alves ${}^1$ and M. S. Kugeratski ${}^2$}
\affiliation{${}^1$
Instituto de F\'{\i}sica e Matem\'atica,  Universidade
Federal de Pelotas, 
Caixa Postal 354, CEP 96010-900, Pelotas, RS, Brazil}
\affiliation{${}^2$ Universidade Federal de Santa Catarina,
Campus Joinville,
Rua Presidente Prudente de Moraes, 406, CEP 89218-000, Joinville, SC, Brazil.}
\begin{abstract}
The description of the QCD dynamics in the kinematical range which will be
probed in the future electron - hadron colliders is still an open question.
Although phenomenological studies indicate that  the gluon number fluctuations,
which are related to discreteness in the QCD evolution, are negligible at HERA,
the magnitude of these effects for the next generation of colliders still should
be estimated. In this paper we investigate inclusive and diffractive $ep$
observables considering a model for the physical scattering amplitude which
describes the HERA data. Moreover, we estimate, for the first time, the
contribution of the fluctuation effects for the nuclear structure functions.
Our results indicate that the
study of these observables in the future colliders can be useful
to constrain the presence of gluon number fluctuations.

\end{abstract}
%
%
\maketitle
%
%
\section{Introduction} \label{sec:intro}
In Quantum Chromodynamics (QCD) at high energies, fluctuation effects arise when
pomeron loop equations are used to describe dipole evolution with increasing
rapidity (see \cite{Hatta06} and references therein). These equations correspond to a generalization of
Balitsky-JIMWLK equations \cite{JKLW,JKLW1,JKLW2,JKLW3,CGC,CGC1,CGC2,W,B} and predict, in the fixed strong coupling case,
the emergence of the diffusive scaling \cite{Hatta06,MS04}. At sufficiently high
energies, this new type of scaling washes out geometric scaling,
a property predicted by the effective theory of Color Glass Condensate
(CGC) \cite{gscaling} and observed in some experiments \cite{SGBK01,GPSS07}.
So far, only few phenomenological analyses looking for diffusive scaling
behavior have been done. Inclusive \cite{Kozlov07,AGBSflu,Xiang09} and
diffractive \cite{Xiang10-ddis} electron-proton deep inelastic scattering
(DIS) at HERA have been investigated, the results indicating no evidence
of fluctuation effects. These have also been studied in the analysis of the
pseudo-rapidity distribution of hadron multiplicities of high energy Au+Au
collisions at RHIC and in predictions for these observables in Pb+Pb
collisions by using Color Glass Condensate dynamics at LHC/ALICE \cite{Xiang10}.
It has been found that the charged hadron multiplicities at central rapidity
are significantly smaller than saturation based calculations and are compatible
to those obtained on a study of multiplicities in the fragmentation region
with running coupling corrections \cite{Albacete07}. Finally, fluctuations
have been investigated in $\gamma^{(*)}\gamma^{(*)}$ collisions at LEP and future $e^+ e^-$ colliders
\cite{GS12} and it has been found that although observing the presence of
the fluctuation effects can be a hard task, they should not be disregarded
in the description of some observables in future colliders.

Although fluctuations have been discarded by toy models which reproduce
some of the main features of high energy evolution and scattering in QCD
\cite{DIPS07,GOS12}, in particular when running coupling corrections
are included, it still lacks a similar treatment in real QCD. Then,
it still seems to be important to look for fluctuations in different
processes and experiments. In this way, and due to the fact that new
electron-hadron colliders have been proposed--the Large Hadron Electron
Collider (LHeC) at CERN \cite{LHeC13} and the Electron Ion Collider (EIC) at
RHIC\cite{EIC13}--in this paper we estimate the contribution of theses effects for inclusive and diffractive observables which will be probed in these future colliders.

The paper is organized as follows: in Section II the expressions
for the observables of interest are presented. In Section III
the procedure of including fluctuations in $ep$ and $eA$ DIS
is described in details. Section IV is devoted to the results of
the phenomenological analysis, as well as predictions in both
processes, and the main conclusions are presented in Section V.

\section{DIS in the dipole frame}\label{sec:dis}

The photon-hadron interaction at high energy (small $x$) is usually described in the infinite momentum frame  of the hadron in terms of the scattering of the photon off a sea quark, which is typically emitted  by the small-$x$ gluons in the proton. However, in order to describe inclusive and diffractive interactions and disentangle the small-$x$ dynamics of the hadron wavefunction, it is more convenient to consider the photon-hadron scattering in the {\it dipole frame}, in which most of the energy is carried by the hadron, while the  photon has enough energy to dissociate into a quark-antiquark pair, before the scattering. The probing projectile fluctuates into a quark-antiquark ($q\bar q$) pair, a \textit{dipole}, with transverse separation $\rr$ long before the interaction, which then
scatters off the target \cite{dipole}. The main motivation to use this color dipole approach is that it gives a simple unified picture of inclusive and diffractive processes.  In this particular frame, the DIS total cross section factorizes and can be written as
   \begin{eqnarray}\label{eq:cross}
      \sigma_{T,L}(x,Q^2) = \int dz \, d^2r \, |\psi _{T,L}(z,\rr,Q^2)|^2 \, \sigma_{dh}(\rr,x) \,\,\,,
   \end{eqnarray}
where $Q^2$ is the photon virtuality, $z$ ($1-z$) is the momentum fraction of the photon carried by the quark
(antiquark) of the dipole and $\rr$ is the transverse size of the dipole. $\psi _{T \,(L)}(z,\rr,Q^2)$
is the wave function which describes the splitting of a transverse (longitudinal) photon into
the dipole, and is known from QED. Moreover, $\sigma_{dh}$ is the dipole-hadron cross section,
which is determined by the QCD dynamics, which will be discussed in the next section. The structure functions read
\begin{equation}\label{eq:f2}
    F_2(x,Q^2)=\frac{Q^2}{4\pi^2\alpha}[\sigma_T+\sigma_L] \,\,\, \mbox{and} \,\,\, F_L(x,Q^2)=\frac{Q^2}{4\pi^2\alpha}\sigma_L \,\,\,,
\end{equation}
where  $\alpha$ is the electromagnetic coupling constant.

In  diffractive DIS events, described by the reaction
$\gamma^*h\rightarrow Xh$, the final states contain an
intact scattered hadron $h$ and a diffractive hadronic state $X$ separated by a
\textit{rapidity gap} $Y_{\textrm{gap}}\equiv \ln(1/x_\mathbb P)$, where
$x_\mathbb P = x /\beta$ and $\beta$ is related to the diffractive invariant mass
$M_X^2$ by
$\beta\equiv Q^2/(Q^2+M_X^2)$. The total diffractive cross sections take the following form in the dipole frame  (See e.g. Ref. \cite{GBW}),
\begin{equation}\label{eq:sigdiff}
\sigma^\mathcal{D}_{T,L} = \int_{-\infty}^0 dt\,e^{B_D t} \left. \frac{d \sigma ^\mathcal{D} _{T,L}}{d t} \right|_{t = 0} = \frac{1}{B_D} \left. \frac{d \sigma ^\mathcal{D} _{T,L}}{d t} \right|_{t = 0}
\end{equation}
where
\begin{equation}\label{eq:dsig-dt}
\left. \frac{d \sigma ^\mathcal{D} _{T,L}}{d t} \right|_{t = 0} = \frac{1}{16 \pi} \int d^2 {\bf r}
\int ^1 _0 d \alpha |\Psi _{T,L} (\alpha, {\bf r})|^2 \sigma _{dh} ^2 (x, \rr).
\end{equation}
It is assumed that the dependence on the momentum transfer, $t$, factorizes and is given by an exponential with diffractive slope $B_D$.
The diffractive DIS (DDIS) can be analysed in detail by studying the behavior
of the diffractive structure function $F_2^{D (3)}(Q^{2}, \beta, x_{I\!\!P})$.
Following Ref. \cite{GBW2}, it is assumed that this structure function is given by
\begin{equation}
F_2^{D (3)} (Q^{2}, \beta, x_{I\!\!P}) = F^{D}_{q\bar{q},L} + F^{D}_{q\bar{q},T} + F^{D}_{q\bar{q}g,T}
\label{soma}
\end{equation}
where $T$ and $L$ again refer to the polarization of the virtual photon, the first and
second terms in the r.h.s. refer to the quark-antiquark ($q\bar{q}$) contribution
and the third one refer to the quark-antiquark-gluon ($q\bar{q}g$)
contribution to DDIS. For the latter, only the transverse polarization is considered,
since the longitudinal counterpart has no leading logarithm in $Q^2$. The $q\bar{q}$
contributions read \cite{fss}
\begin{equation}
  x_{I\!\!P}F^{D}_{q\bar{q},L}(Q^{2}, \beta, x_{I\!\!P})=
\frac{3 Q^{6}}{32 \pi^{4} \beta B_D} \sum_{f} e_{f}^{2}
 2\int_{\alpha_{0}}^{1/2} d\alpha \alpha^{3}(1-\alpha)^{3} \Phi_{0},
\label{qqbl}
\end{equation}
\begin{equation}
 x_{I\!\!P}F^{D}_{q\bar{q},T}(Q^{2}, \beta, x_{I\!\!P}) =
 \frac{3 Q^{4}}{128\pi^{4} \beta B_D}  \sum_{f} e_{f}^{2}
 2\int_{\alpha_{0}}^{1/2} d\alpha \alpha(1-\alpha)
\left\{ \epsilon^{2}[\alpha^{2} + (1-\alpha)^{2}] \Phi_{1} + m_f^{2} \Phi_{0}  \right\}
\label{qqbt}
\end{equation}
where the lower limit of the integral over $\alpha$ is given by
$\alpha_{0} = \frac{1}{2} \, \left(1 - \sqrt{1 - \frac{4m_{f}^{2}}{M_X^{2}}}\right)$
and
\begin{equation}
\Phi_{0,1}  \equiv  \left(\int_{0}^{\infty}r dr K_{0 ,1}(\epsilon r)\sigma_{dh}(x_{I\!\!P},r) J_{0 ,1}(kr) \right)^2.
\label{fi}
\end{equation}
The $q\bar{q}g$ contribution within the dipole picture at leading $\ln Q^2$ accuracy
is given by \cite{fss}
 \begin{eqnarray}
   \lefteqn{x_{I\!\!P}F^{D}_{q\bar{q}g,T}(Q^{2}, \beta, x_{I\!\!P})
  =  \frac{81 \beta \alpha_{s} }{512 \pi^{5} B_D} \sum_{f} e_{f}^{2}
 \int_{\beta}^{1}\frac{\mbox{d}z}{(1 - z)^{3}}
 \left[ \left(1- \frac{\beta}{z}\right)^{2} +  \left(\frac{\beta}{z}\right)^{2} \right] } \label{qqg} \\
  & \times & \int_{0}^{(1-z)Q^{2}}\mbox{d} k_{t}^{2} \ln \left(\frac{(1-z)Q^{2}}{k_{t}^{2}}\right)
\left[ \int_{0}^{\infty} u \mbox{d}u \; \sigma_{dh}(u / k_{t}, x_{I\!\!P})
   K_{2}\left( \sqrt{\frac{z}{1-z} u^{2}}\right)  J_{2}(u) \right]^{2}.\nonumber
\end{eqnarray}
As pointed in Ref. \cite{marquet}, at small $\beta$ and low $Q^2$, the leading $\ln (1/\beta)$ terms should be resummed and the above expression should be modified. However, as a description with the same quality using the Eq. (\ref{qqg}) is possible by adjusting the coupling $\alpha_s$ \cite{marquet}, in what follows we will use this expression for our phenomenological studies.

\section{QCD dynamics}

Let us consider the problem of a scattering between a small dipole
(a colorless quark-antiquark pair) and a dense hadron target, at a given
rapidity interval $Y=\ln(1/x)$. The dipole has transverse size given by the vector
$\rr=\xx-\yy$, where $\xx$ and $\yy$ are the transverse vectors for the quark
and antiquark, respectively, and impact parameter $\bb=(\xx+\yy)/2$. The  dipole-hadron cross section,
$\sigma_{dh}$, can be expressed as
    \begin{equation}\label{eq:dipcross}
        \sigma_{dh}(\rr,x)=2\int d^2b\,{\cal{N}}_h(\bb,\rr,x),
    \end{equation}
where ${\cal{N}}_h(\bb,\rr,x)$ is the imaginary part of the forward amplitude of the
dipole-hadron scattering, at a given impact parameter $\bb$ and a rapidity interval $Y=\ln(1/x)$.
This quantity encodes all the information about the hadronic scattering, and thus about the
non-linear and quantum effects in the hadron wave function.
The evolution with the rapidity $Y$ of
${\cal{N}}_h(\rr,\bb,Y)\equiv {\cal{N}}_h(\xx,\yy,Y)\equiv\calN_Y(\xx,\yy)$  is given by an infinite hierarchy of equations, the so called
Balitsky-JIMWLK equations \cite{JKLW,JKLW1,JKLW2,JKLW3,CGC,CGC1,CGC2,W,B}.
In the mean field approximation, this infinite set of coupled equations reduces to
a single one, the Balitsky-Kovchegov (BK) equation \cite{B,K}, a closed equation
for the one-dipole scattering amplitude $\calN_Y(\xx,\yy)$, which, at fixed coupling,
is given by
\begin{equation}\label{eq:bk}
\partial_Y \calN_Y(\xx,\yy) = \bar{\alpha}\int d^2z\,
\frac{(\xx-\yy)^2}{(\xx-\zz)^2(\zz-\yy)^2}
\left[\calN_Y(\xx,\zz)+\calN_Y(\zz,\yy)-\calN_Y(\xx,\yy)
-\calN_Y(\xx,\zz)\calN_Y(\zz,\yy)\right],
\end{equation}
where $\bar{\alpha}=\alpha_sN_c/\pi$. In the translation invariance approximation, the amplitude is independent of the impact parameter $\bb$ and depends only on the dipole size $r=|\rr|$, i.e. $\calN_Y(\rr)=\calN_Y(r)$.
For small values of $r$, the BK solution $\calN_Y(r)$ is small -- the color transparency regime -- and the linear solution is enough to describe the dipole evolution. For large $r$, the amplitude approaches the unitarity bound, or 'black disk' limit $\calN(\rr)=1$. The transition between these two regimes takes place at $r=1/Q_s(Y)$, where $Q_s(Y)$ is an increasing function of rapidity $Y$ and is called the {\it saturation scale}, defined in such a way that $\calN(\rr)={\cal O}(1)$ when $r=1/Q_s(Y)$.

The BK equation admits travelling wave solutions
\cite{MP03}: at asymptotic rapidities, the scattering amplitude depends only on the ratio $r^2Q_s^2(Y)$ instead of depending separately on $r$ and $Y$. This scaling property is called geometric scaling and has been observed in the measurements of the proton structure function at HERA \cite{gscaling}. The amplitude is a wavefront which interpolates between 0 and 1 and travels towards smaller values of $r^2$ with speed $\lambda$ -- the saturation exponent -- keeping its shape, and the saturation scale $Q_s(Y)$ gives the front position.

Besides the theoretical explanation of geometric scaling in terms of travelling wave solutions of
BK equation, the correspondence between QCD evolution at high energy and reaction-diffusion
processes has also brought to light the fact that the Balitsky-JIMWLK hierarchy misses important effects,
those due to gluon (dipole) number fluctuations, which are related to discreteness in the evolution
\cite{IMM05}. At least at fixed coupling \cite{IMM05,Hatta06,onedim,BT13}, fluctuations influence dramatically the QCD evolution at high energies and, when they are included, a new hierarchy of evolution equations arise,
the {\it pomeron loop} equations \cite{IT05}. These are rather complicated to solve, so
some approximations are needed to get some knowledge about the dipole scattering amplitudes. After
such approximations \cite{IT05}, the new hierarchy has been found to can be generated from a Langevin equation for the single-event amplitude, which formally is the BK equation with a noise term, which lies in the same universality class of the stochastic FKPP equation (sFKPP): each realization of the noise means a single realization of the target in the evolution and leads to an amplitude for a single event. Different realizations of the target lead to a dispersion of the solutions, and then in the saturation momentum $\rho_s\equiv \ln(Q_s^2/Q_0^2)$
from one event to another. The saturation scale is now a random variable whose average value is given by
\begin{equation}
\langle Q_s^2(Y) \rangle = \exp{[\lambda^*Y]}
\end{equation}
and the dispersion in the position of the individual fronts is given by
\begin{equation}
\sigma^2 = \langle \rho_s^2 \rangle - \langle \rho_s \rangle^2 = D\bar{\alpha}Y.
\end{equation}
where $D$ is the \textit{diffusion coefficient}, a number expected to be of
order one, which determines the rapidity $Y_D = 1/D$ above which gluon number fluctuations become important.

The probability distribution of $\rho_s$ is, to a good approximation, a Gaussian \cite{MS06}
\begin{equation}
P_Y(\rho_s)\simeq \frac{1}{\sqrt{\pi\sigma^2}}\exp\left[-\frac{(\rho_s-\avg{\rho_s})^2}{\sigma^2}\right].
\end{equation}
The travelling-wave behavior is kept by the evolved single event amplitude, that is,
geometric scaling is preserved for each realization of the noise (correspondingly, each
realization of the target configuration at rapidity $Y$). However, the speed $\lambda^*$ of the wave is smaller
than that predicted by BK equation. The physical (average over all the configurations of the target)
amplitude is obtained through ($\rho\equiv \ln(1/r^2Q_0^2)$)
\begin{equation}\label{eq:avg-amplitude}
\avg{{\cal{N}}_h(\rho,\rho_s)} = \int^{+\infty}_{-\infty}d\rho_s\,P_Y(\rho_s){\cal{N}}_h(\rho,\rho_s)\,\,,
\end{equation}
where ${\cal{N}}_h(\rho,\rho_s)$  is the event-by-event scattering amplitude. Now, at sufficiently high energies, unlike the individual fronts, the amplitudes will generally not show geometric scaling, but will show additional dependencies upon $Y$, through the front dispersion $\sigma$. As a consequence, geometric scaling is washed out and replaced by the so-called \textit{diffusive scaling} \cite{IMM05,Hatta06}
\begin{equation}
\avg{{\cal{N}}_h(\rho,\rho_s)} = {\cal{N}}\left(\frac{\rho-\avg{\rho_s}}{\sqrt{\bar{\alpha}DY}}\right),
\end{equation}
that is, at asymptotic rapidities, the average amplitude will depend on the diffusive scaling variable
$(\rho-\avg{\rho_s})/\sqrt{\bar{\alpha}DY}$.

\section{Results}\label{sec:results}

The search for evidences of the gluon number fluctuation effects has motivated the analysis performed in Refs.   \cite{Kozlov07,AGBSflu,Xiang09,Xiang10-ddis}, which have looked for any signal of the diffusive scaling in the kinematical region probed by HERA. The results of these studies indicate no evidence
of fluctuation effects. However, recent theoretical
developments \cite{BT13} indicate that, at least at fixed coupling (and at the level of single pomeron loop), fluctuations should be taken into account  at the energies probed by the future electron - hadron colliders. This motivates us to extend the previous studies to the  kinematical range which could be probed in these colliders \cite{LHeC13,EIC13} and estimate, for the first time, the magnitude of the fluctuation effects for the small-$x$ behavior of the nuclear structure functions. In what follows we estimate the observables considering as input in the calculations models for the physical scattering amplitude $\avg{{\cal{N}}_h}$ and compare our predictions with those obtained using the event-by-event scattering amplitude ${\cal{N}}_h$, i.e. disregarding the fluctuaction effects.

\begin{figure}[t]
\scalebox{0.5}{\includegraphics{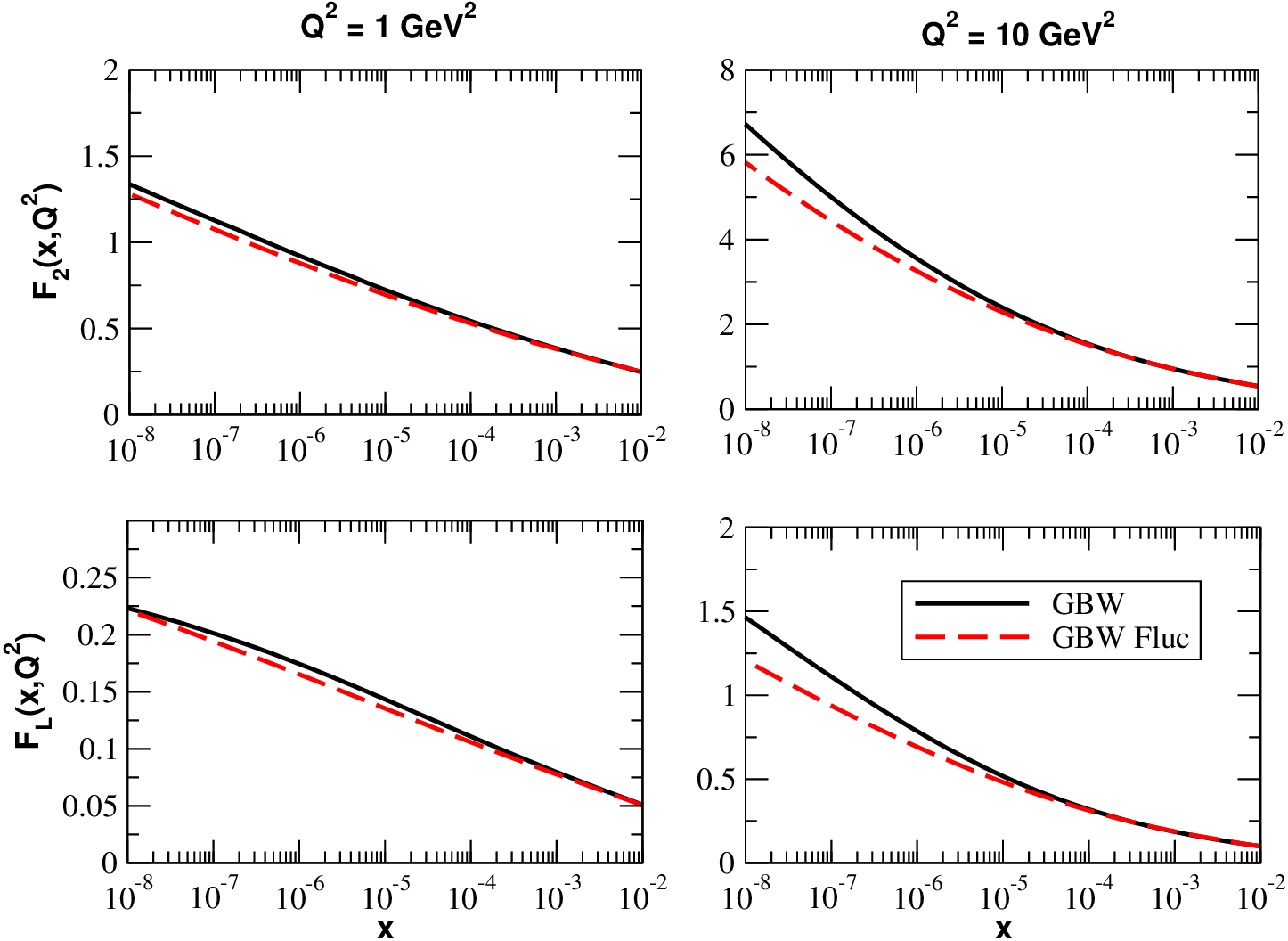}}
\caption{(Color online) Proton structure functions as functions of $x$ for different values of $Q^2$.}
\label{fig:f2p}
\end{figure}

\subsection{Electron - proton collisions}

{ Let us begin our analysis discussing $ep$ collisions, which could be studied in the future LHeC collider. The starting point
for evaluating} the inclusive and diffractive observables described in the Section \ref{sec:dis},
{ is to specify  the single event amplitude ${\cal{N}}_p$. Following Refs.  \cite{GBW,Kozlov07}, we will use, for ${\cal{N}}_p$, the GBW model}
\begin{equation}\label{eq:gbw}
{\cal{N}}_p(r,Y)=1-e^{-r^2Q_s^2(Y)/4},
\end{equation}
where the saturation scale is given by $Q_s^2(Y\equiv\ln(x_0/x))=Q_0^2\left(x_0/x\right)^{\lambda}$,
$x_0$ is the value of the Bjorken $x$ in the beginning of the evolution and $\lambda$ is the
saturation exponent. { We assume} the translational invariance approximation, which regards
hadron homogeneity in the transverse plane, which implies that the dipole-proton cross section and the forward dipole scattering amplitude are related by a constant $\sigma_0$, which results from the $\bb$ integration and sets the normalization. In Ref. \cite{Kozlov07} the authors found that the description of the DIS data is improved once gluon number fluctuations are included and that the values of the saturation exponent and the diffusion coefficient turn out reasonable and agree with values obtained from numerical simulations of toy models which take into account fluctuations.  For instance, for the event-by-event amplitude given by Eq. (\ref{eq:gbw}),  they have found that $\lambda = 0.225$, $x_0 = 0.0546 \times 10^{-4}$ and $D = 0.397$ for a $\chi^2/$d.o.f. = 1.14. In contrast, for the $D = 0$ case (no fluctuations), $\lambda = 0.285$, $x_0 = 4.11 \times 10^{-4}$  and $\chi^2/$d.o.f. = 1.74.
A similar conclusion was obtained considering the IIM model \cite{IIM03} for the event-by-event amplitude, with the values of $\lambda$ and $D$ being quite model independent. As demonstrated in Ref. \cite{GS12}, when the gluon fluctuations effects are included, the onset of saturation is strongly delayed in comparison to the event-by-event scattering amplitude. This motivates us to quantify the magnitude of the gluon number fluctuations effects for the energy of the future colliders. As in \cite{Kozlov07} we assume that $\sigma_0 = 2 \pi R^2 $, with $R$ being a free parameter fitted by the HERA data which is equal to 0.594 (0.712) fm for the event-by-event (physical) amplitude.  It is important to emphasize that our predictions for inclusive observables for the kinematical range probed by future colliders are parameter free, since all parameters have been fixed by the HERA data.

\begin{figure}[t]
\begin{tabular}{cc}
\includegraphics[scale=0.25]{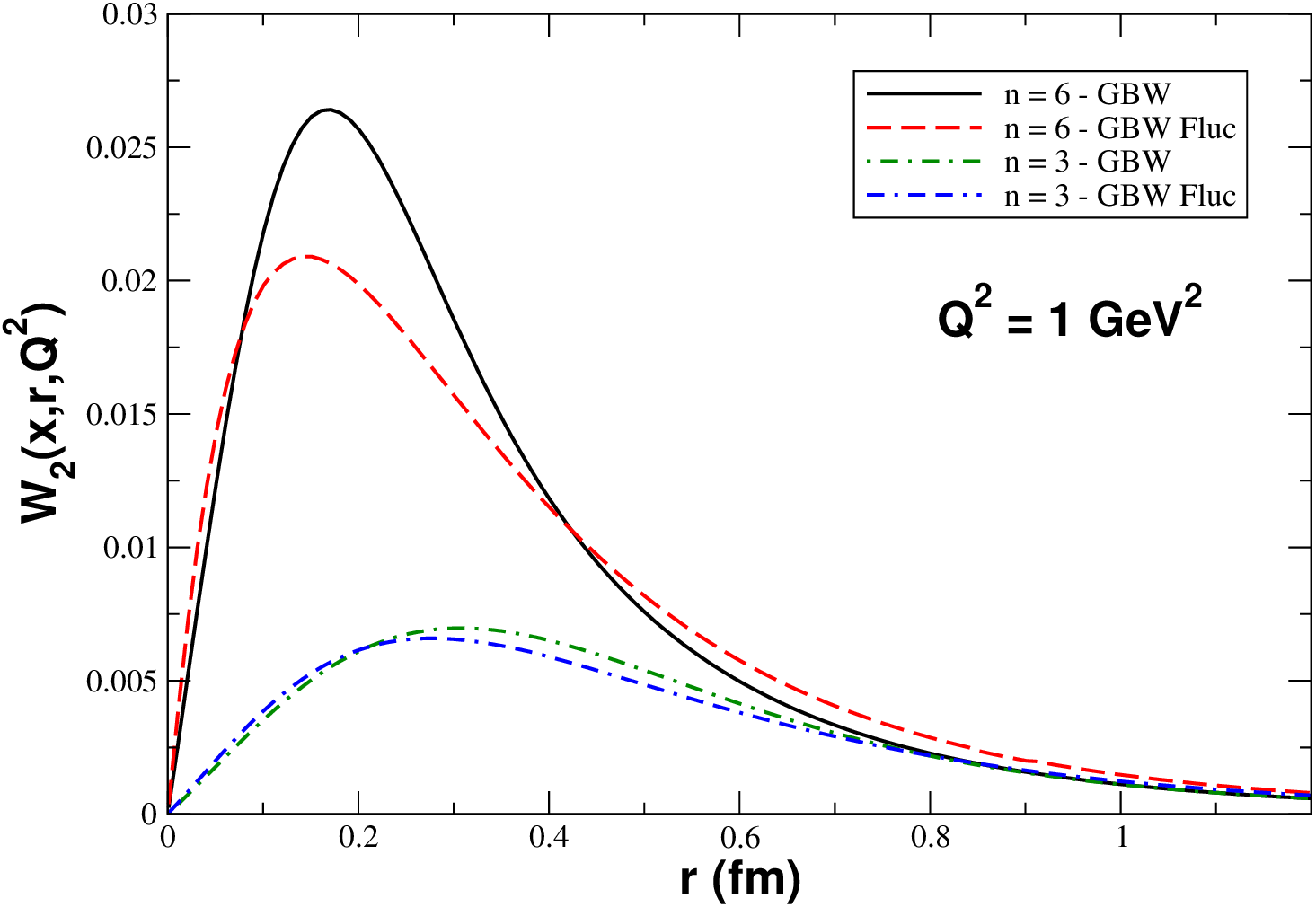} & \includegraphics[scale=0.25]{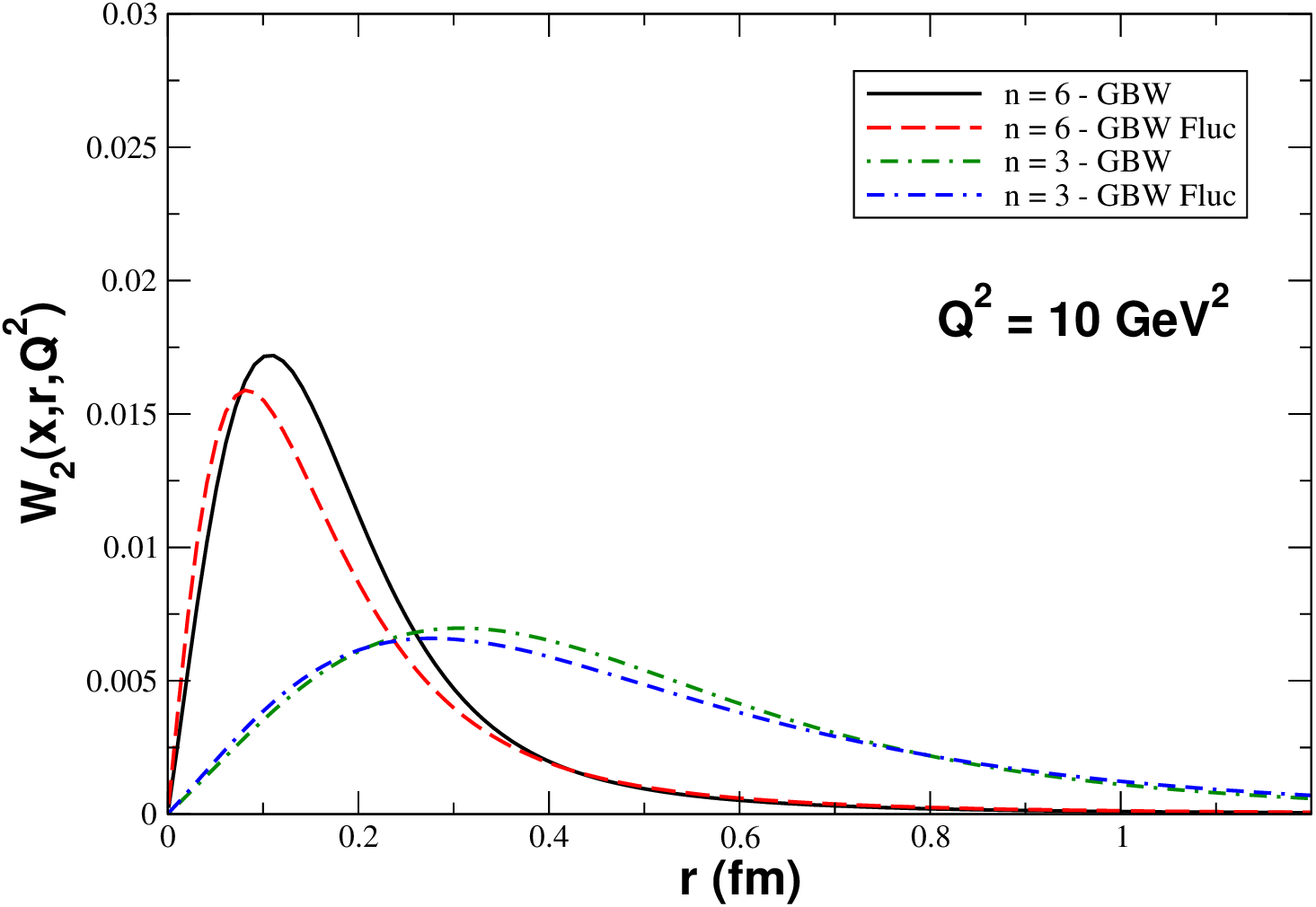}\\
\includegraphics[scale=0.25]{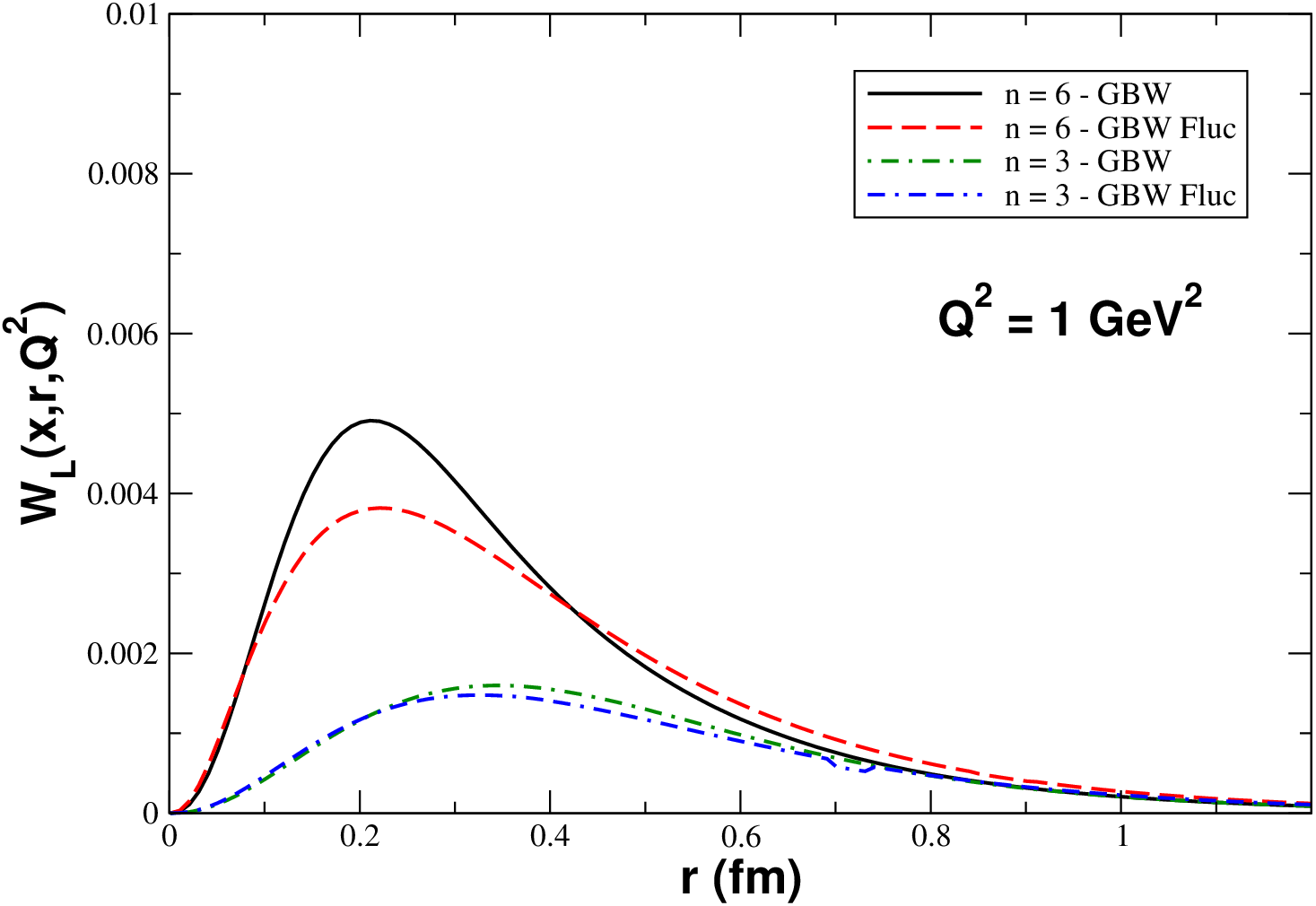} & \includegraphics[scale=0.25]{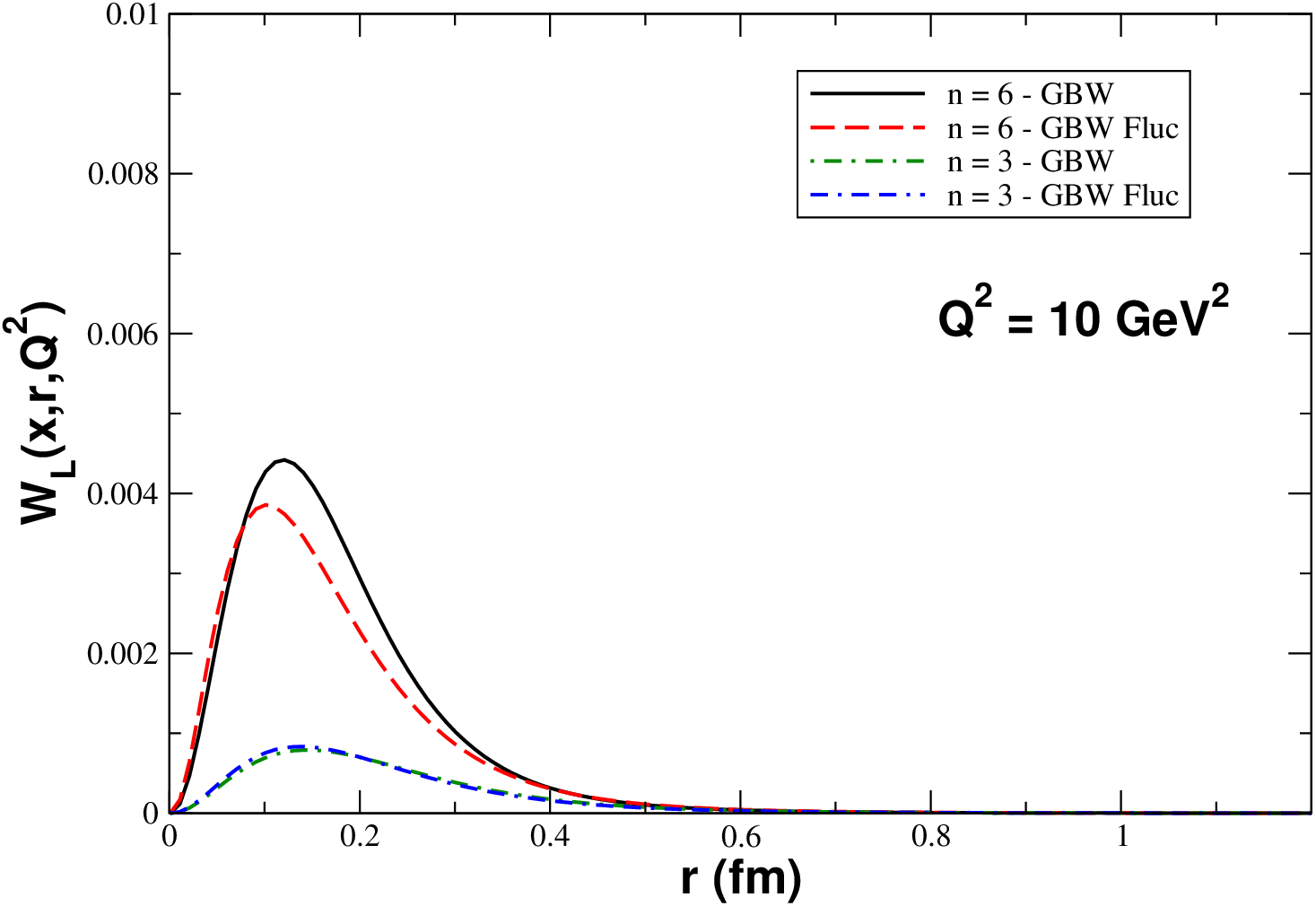}
\end{tabular}
\caption{ (Color online) The {\bf r}-dependence of the photon-nucleon overlap functions for different values the small-$x$ ($x=10^{-n}$) and $Q^2$. }
\label{overlap}
\end{figure}

In Fig. \ref{fig:f2p} we present our predictions for the proton structure functions $F_2$ and $F_L$ considering two characteristic values of the photon virtuality $Q^2$. We can see that at small values of $Q^2$, the predictions with fluctuation (denoted GBW Fluc) and without fluctuation (denoted GBW) are almost identical. In contrast, at $Q^2 = 10$ GeV$^2$, we predict a difference  $\le  20$ (7.0)  \%  for $F_L$ ($F_2$). This behavior is expected since the gluon number fluctuations  modify the transition between the linear and nonlinear regimes and should be more important at smaller dipoles, which are probed at larger values of $Q^2$.  The contribution of small and large pair separations for inclusive observables can be studied in more detail considering the corresponding overlap functions which are given by:
\begin{eqnarray}
W_{2} \,(r,x,Q^2)  = { 2\pi r} \sum_{i=T,L} \, \int dz\, |\Psi_i
(z,\,r, Q^2)|^2 \, \sigma_{dp} (x,r)\, ,
\label{eq:overlap-incl}
\end{eqnarray}
and
\begin{eqnarray}
W_{L} \,(r,x,Q^2)  = { 2\pi r}  \, \int dz\, |\Psi_L
(z,\,r, Q^2)|^2 \, \sigma_{dp} (x,r)\, .
\label{eq:overlap-long}
\end{eqnarray}
It is important to emphasize that the behavior of the overlap function is strongly dependent on the scattering amplitude.  Moreover, it is the energy dependence of the scattering amplitudes which determine the  $x$ dependence of the overlap function.  In Fig. \ref{overlap} we present our predictions for the overlap functions for different values of $x$ and $Q^2$ considering the event-by-event and the physical amplitudes. As anticipated, the overlap functions peak at smaller values of $r^2$ at larger values of $Q^2$. We have that at  $x = 10^{-3}$ the predictions are very similar. At $x = 10^{-6}$ the GBW Fluc prediction, which takes into account the gluon number fluctuations, is smaller than  the GBW one in almost the full range of $r^2$, which explains the behavior observed in Fig. \ref{fig:f2p}.

\begin{figure}[t]
\scalebox{0.5}{\includegraphics{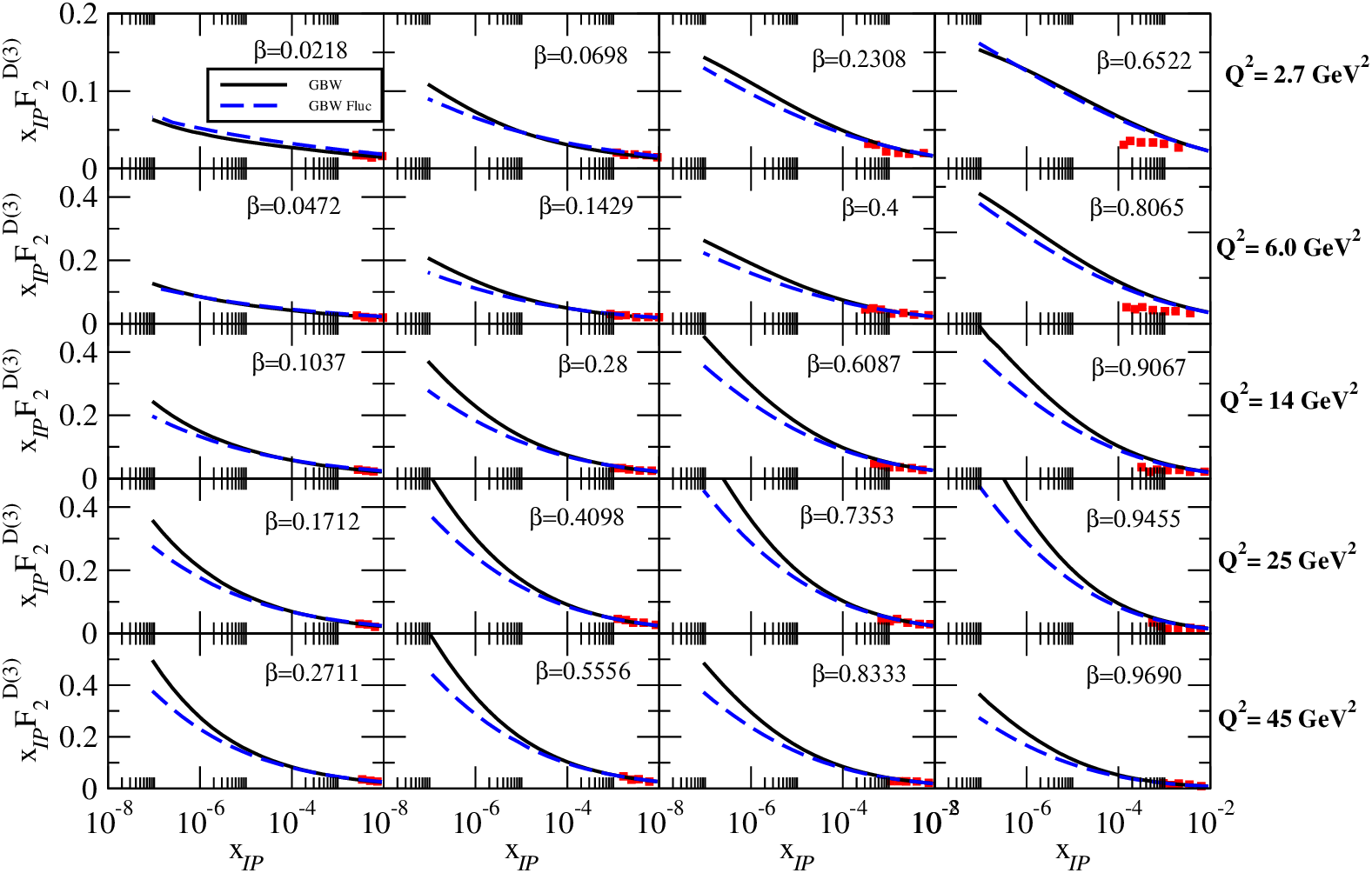}}
\caption{(Color online)  Predictions for $F_2^{D (3)} (Q^{2}, \beta, x_{I\!\!P})$
compared with the ZEUS data \cite{zeusdata}, using GBW (solid line) and GBW Fluc (dashed line) models.}
\label{fig:f2d3}
\end{figure}

In order to calculate the diffractive structure function and compare with the HERA data we need to specify the diffractive slope $B_D$ and the coupling $\alpha_s$, which determine the normalization of $F_2^{D (3)} (Q^{2}, \beta, x_{I\!\!P})$. In particular, the magnitude of the $q\bar{q}g$ contribution is strongly dependent on the value of $\alpha_s$. In our calculations we choose $B_D = 7.3$ GeV$^{-2}$, which is  in reasonable agreement with the experimental data \cite{aktas}. On the other hand, we are still free to choose the value of $\alpha_s$. Following \cite{fss}, we assume $\alpha_s = 0.15$. In a more detailed study we could consider its running with $Q^2$ or perform a fit to experimental data. However, as our goal is to estimate the magnitude of the gluon number fluctuations effects, we postpone this study to a future publication.
The diffractive cross section $ep \rightarrow eXY$ have been measured by the H1 and ZEUS experiments at HERA tagging the proton in the final state ($Y = p$) or selecting events with a large rapidity gap between the systems $X$ and $Y$ in the case of H1  and using the $M_X$-method in case of ZEUS. The distinct methods and experimental cuts used by the H1 and ZEUS collaborations imply data with differences in the normalization. Moreover, while the ZEUS data are given for the diffractive structure function $F_2^{D (3)}$, the H1 ones are presented for the reduced cross section which is expressed in terms of a combination of diffractive structure functions and kinematical factors. As our predictions are for $F_2^{D(3)}$, as given in Eq. (\ref{soma}), we restrict our comparison to the recent  ZEUS data \cite{zeusdata}. Furthermore, as the dipole model is more suitable for the description of the diffractive structure functions in the region of low and moderate $Q^2$, we restrict our comparison to the experimental data in the kinematical region of $Q^2 < 50$ GeV$^2$ and $x_{I\!\!P} \le 10^{-2}$. In Fig. \ref{fig:f2d3}
we compare our predictions  with the
ZEUS data \cite{zeusdata}  for $F_2^{D(3)}$ for five values of the photon virtuality $Q^2$.
It can be seen that both models describe reasonably the experimental data, with the predictions being almost identical in the kinematical range probed by HERA. However, they differ at smaller values of $x_{I\!\!P}$, in particular by  $\approx 35$ \%  at larger values of $\beta$ and $Q^2$.

\begin{figure}[t]
\scalebox{0.4}{\includegraphics{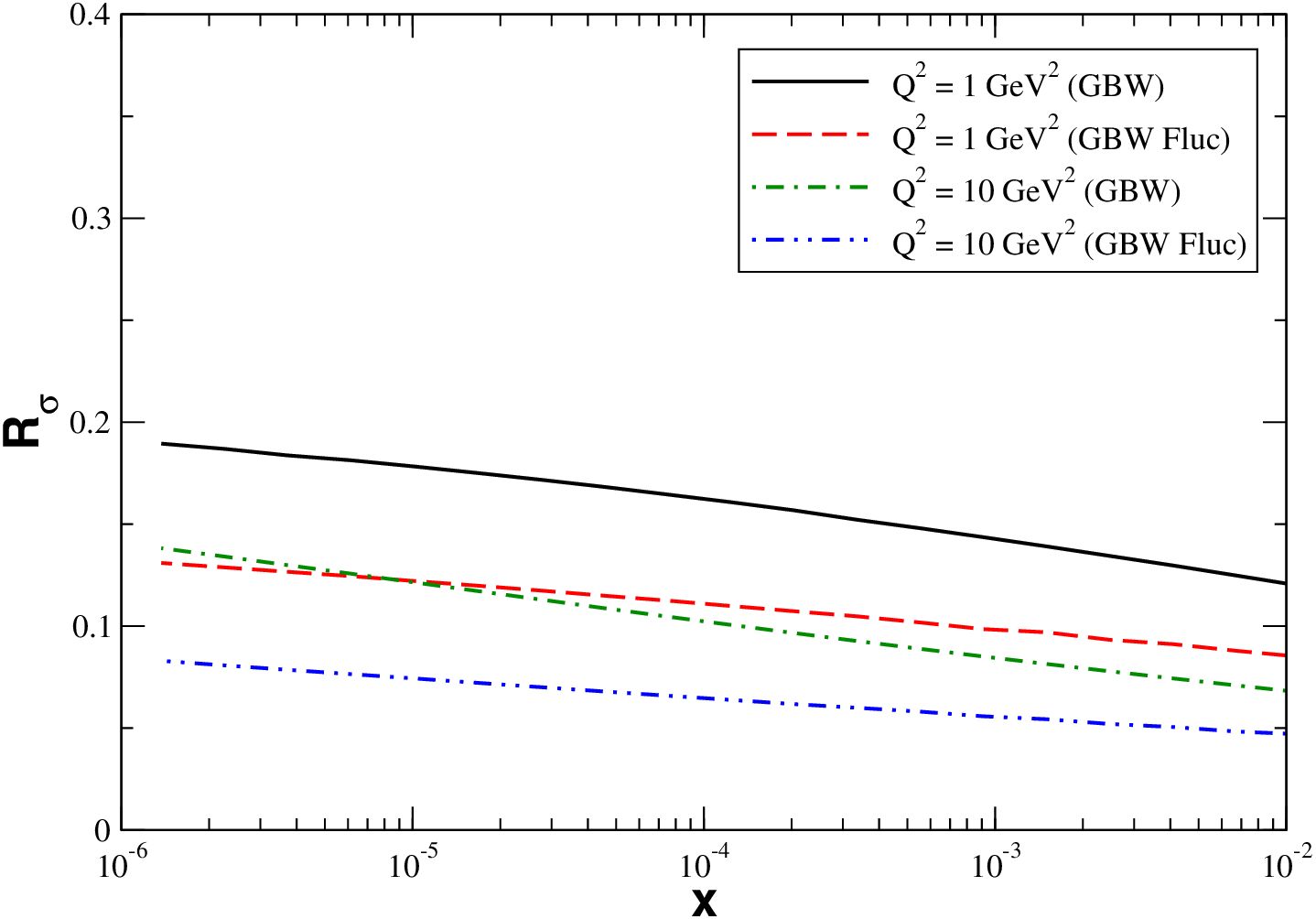}}
\caption{(Color online)  Ratio of  the diffractive to total cross sections, $R_{\sigma}= \sigma_{diff}/\sigma_{tot}$, as a function of $x$.}
\label{fig:ratio-diff-tot}
\end{figure}

Another observable of interest is the ratio between the diffractive and total cross sections, $R_{\sigma} = \sigma_{diff}/\sigma_{tot}$, which has been measured in $ep$ diffractive scattering by the ZEUS collaboration \cite{cheka}. Experimentally, it is observed a very similar energy dependence of the inclusive diffractive
and the total cross section, with the saturation models providing a simple explanation for this finding \cite{GBW}. In Fig. \ref{fig:ratio-diff-tot} we present our predictions for $R_{\sigma}$ as a function of $x$ and different values of $Q^2$. It can be seen that the fluctuations effects reduces the ratio by $\approx$ 40 \% at very small-$x$.

\subsection{Electron - nucleus collisions}

Let us now consider the influence of gluon number fluctuations in the nuclear
structure functions which could be probed in the future electron-ion colliders. As in the $ep$ case, the main input in our calculations using the color dipole approach is { now} the forward dipole-nucleus scattering amplitude ${\cal{N}}_A$
or, equivalently, the dipole-nucleus cross section $\sigma_{dA}$.  The description of  ${\cal{N}}_A$ is still an open question in the literature and it still lacks a deep theoretical study. Here,
we will  consider three different phenomenological models, which are based on distinct assumptions and allow us to obtain an estimate of the contribution of the fluctuation effects and theoretical uncertainties present in the predictions for the nuclear case.

The first model we will use for the nuclear event-to-event scattering amplitude (called Model I hereafter) was first proposed in Ref. \cite{armesto_epjc}
and has shown to successfully describe the data on the nuclear structure function, $F_2^A$.
It provides the following expression for the event-to-event dipole-nucleus  scattering amplitude
\begin{equation}\label{eq:NA}
    {\cal{N}}_A(r,x,b)=1 - \exp \left[-\frac{1}{2}A \,T_A(\bb) \, \sigma_0 {\cal{N}}_p(x,\rr^2)\right].
\end{equation}
Here, $T_A(\bb)$ is the nuclear profile function, which is
obtained from a 3-parameter Fermi distribution for the nuclear
density normalized to unity and ${\cal{N}}_p$ is the dipole - proton event-to-event scattering amplitude as given in previous subsection (For a more recent study see Ref. \cite{babi_recent}). The above
equation, based on the Glauber-Gribov formalism \cite{gribov},  sums up all
the multiple elastic rescattering diagrams of the $q \overline{q}$ pair
and is justified for large coherence length, where the transverse separation
$r$ of partons in the multiparton Fock state of the photon becomes a conserved
quantity, {\it i. e.} the size of the pair $r$ becomes eigenvalue
of the scattering matrix. As Eq. (\ref{eq:NA}) represents the classical
limit of the Color Glass Condensate  \cite{raju_acta}, it is  expected to
be modified by quantum corrections at energies larger than those probed
by the current lepton - nucleus  data. The description of   ${\cal{N}}_A$
in the CGC formalism considering these corrections is still an open question.

 In this phenomenological study, it is important to estimate the theoretical
uncertainties present in our calculations. Thus, we will consider a second model
(called Model II hereafter), where the event-by-event dipole - nucleus scattering amplitude  is given by
\begin{equation}\label{eq:NA2}
{\cal{N}}_A (x,\rr) = \left[1-\exp\left(-\frac{\rr^{2}Q^{2}_{s,A}(x)}{4}    \right)  \right].
\end{equation}
which is similar to GBW model used in the proton case,
with the following replacements: $Q^{2}_{s}\to Q^{2}_{s,A}=A^{1/3}Q^{2}_{s}$.
This model implies that the dipole-nucleus cross section can be expressed by
$\sigma_{dA}(r,x) = \sigma_{0,A}{{\cal{N}}_A(r,x)}$, with   $\sigma_{0,A}=A^{2/3}\sigma_{0}$.
The basic assumption of the Model II is that it assumes that the nucleus is so dense that it can be seen as a large hadron with a continuous particle distribution (For details see Ref. \cite{simone1}). Therefore, { it can be considered} as a first approximation for  the asymptotic regime of the saturation physics at very large energies. It must be pointed out that this model does not describe the current experimental data and should be considered as a extreme approach, used here only to estimate the  uncertainty associated to the choice of dipole - nucleus cross section.
These two models for the event-to-event dipole - nucleus scattering amplitude are used as input in the calculation of the physical amplitude
\begin{equation}\label{eq:avg-amplitude-nuc}
\avg{{\cal{N}}_A(\rho,\rho_s)} = \int^{+\infty}_{-\infty}d\rho_s\,P_Y(\rho_s){\cal{N}}_A(\rho,\rho_s)\,\,.
\end{equation}
Furthermore, we consider a third model for $\avg{{\cal{N}}_A}$, denoted Model III hereafter, where we assume that
\begin{equation}
\avg{{\cal{N}_A}} (r,x,b)=1 - \exp \left[-\frac{1}{2}A \,T_A(\bb) \, \sigma_0 \avg{{\cal{N}}_p}(x,\rr^2)\right] \,\,.
\end{equation}
The basic assumption is that in this model we assume that the fluctuaction effects are important in the dipole - proton interaction but can be disregarded at the nuclear level. Like Model II, this model should also be considered an ansatz for the description of the dipole - nucleus interaction in $eA$ collisions for the energies of the future colliders.

\begin{figure}[t]
\scalebox{0.4}{\includegraphics{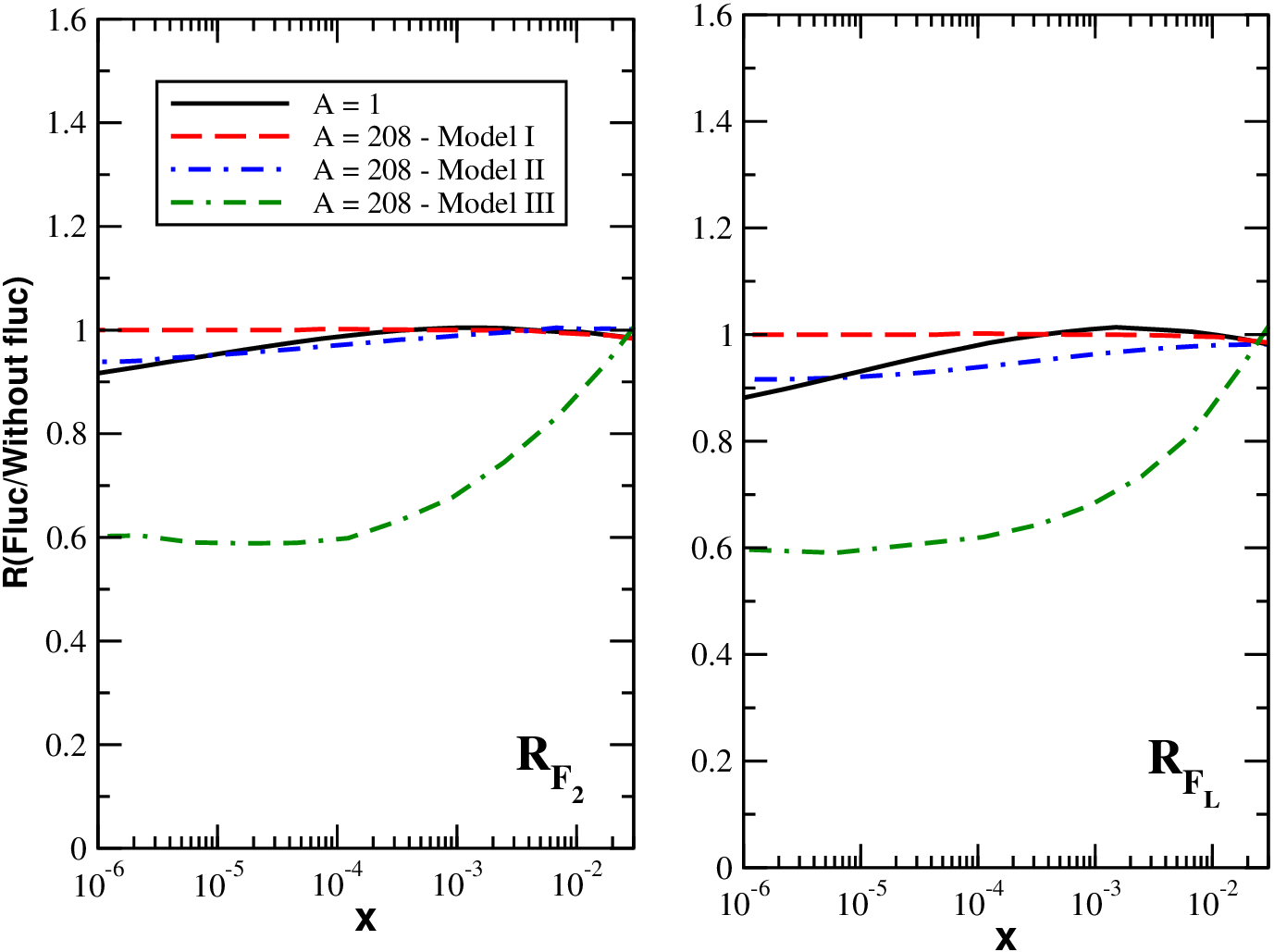}}
\caption{(Color online)  Ratio between predictions with and without fluctuations
for the different nuclear structure functions for $Q^2 = 10$ GeV$^2$. The ratio for the proton (solid line) also is presented for comparison.}
\label{fig:ratio-pA}
\end{figure}

As our goal is to estimate if in electron - ion collisions the magnitude of the gluon number fluctuations
effects is or not amplified, we have quantified the ratio between predictions with and without fluctuations
for the different nuclear structure functions. In Fig. \ref{fig:ratio-pA} we present our results  for $Q^2 = 10$ GeV$^2$ obtained considering the three models discussed above. For comparison we also present the behavior of this ratio  for the proton case.  We can see that the influence of the fluctuation effects for the Model I is very small. On the other hand, for the Model II, the contribution of the fluctuation effects is smaller than 8 \% in the kinematical range considered. Finally, for the Model III, we predict a very large different between the predictions with and without fluctuations effects. As the predictions obtained disregarding the fluctuaction effects describe the scarce available $eA$ data, we expect that a future experimental analysis of the nuclear structure functions could be useful to demonstrate the presence of the fluctuation effects or discard the Model III as a possible model to describe the physical amplitude $\avg{{\cal{N}}_A}$.

\section{Conclusions}

The future electron-hadron colliders should probe  values of  $x$ smaller than HERA and, for a first time, explore this kinematical range with nuclear targets. On the theoretical side, we expect an amplification of the nonlinear effects in the QCD dynamics in this unexplored regime. One of the open questions in the QCD description of the observables is the magnitude of the gluon number fluctuation effects, which are expected to be present when pomeron loop equations are used to describe dipole evolution with increasing rapidity. Aiming at looking for
any evidence of these effects in the next generation of colliders, in this paper we have estimated them in inclusive and diffractive $ep$ observables considering a model which is able to describe the HERA  experimental data. Moreover, we have also extended our study to electron-ion collisions. Our main conclusion is that the experimental analysis of the inclusive  and diffractive structure functions in future electron - hadron colliders can be useful to constrain the presence of gluon number fluctuations.

\begin{acknowledgments}
This work was  partially financed by the Brazilian funding
agencies CNPq, CAPES and FAPERGS.
\end{acknowledgments}

\end{document}